\documentclass[twocolumn,showpacs,preprintnumbers,amsmath,amssymb]{revtex4}

\usepackage{graphicx}
\usepackage{dcolumn}
\usepackage{bm}

\begin{document}

\title{Disorder-Induced First Order Transition and Curie Temperature Lowering   in Ferromagnatic Manganites. }

\author{J. Salafranca and L. Brey}

\affiliation{\centerline { Instituto de Ciencia de Materiales de Madrid
(CSIC),~Cantoblanco,~28049~Madrid,~Spain.}}

\date{\today}

\begin{abstract}
We study the effect that size disorder   in the cations surrounding
manganese ions has on the magnetic properties of manganites. This
disorder is mimic with a proper distribution of spatially disordered
Manganese energies. Both,  the Curie temperature and the order of
the transition are strongly affected by disorder. For moderate
disorder the Curie temperature decreases linearly with the the
variance of the distribution of the manganese site energies, and for
a disorder comparable to that present in real materials the
transition becomes first order. Our results provide a theoretical
framework to understand disorder effects on the magnetic behavior of
manganites.

\end{abstract}

 \maketitle


%

Compounds with chemical formula R$_{1-x}$A$_x$MnO$_3$ (R=trivalent
element and  A= divalent element) are usually called manganites.
These materials exhibit an extraordinarily large (so called
colossal) magnetoresistance. The study of the origin of the colossal
magnetoresistance as well as its possible technological applications
has triggered in the last  years an intensive research on these
materials\cite{dagottobook,tokura_99,coey_99}. This research has
revealed a rich variety of exotic phenomena, and after twenty years
of intense study new and exciting phenomena still appear when
studying manganites\cite{Loudon05,Milward05,brey04,Brey_2005b}.

In manganites the ferromagnetic order is driven by the motion of the
carriers and, therefore, their properties depend on the competition
between kinetic energy, tending to delocalize the carriers, and
localization effects, such as antiferromagnetic coupling between the
Mn core spins and the Jahn-Teller  coupling. Disorder also reduces
the carriers mobility, and, thus, it strongly affects the stability
of the ferromagnetic (FM) phases.
Recently it has been possible to synthesize  cation  ordered
Ln$_{0.5}$Ba$_{0.5}$MnO$_3$, where Ln is a rare earth and LnO and
BaO planes alternates along the $c$ axis\cite{Millange_98}. These
half-doped manganites may give a good benchmark to test theories of
the influence of disorder on the electronic and magnetic properties
of manganites. Experimental work in FM La$_{0.5}$Ba$_{0.5}$MnO$_3$
shows\cite{Sato_04} that disorder strongly suppress the transition
temperature  and changes the character of the transition, form
continuous to weakly first order.

\begin{figure}
  \includegraphics[clip,width=9cm]{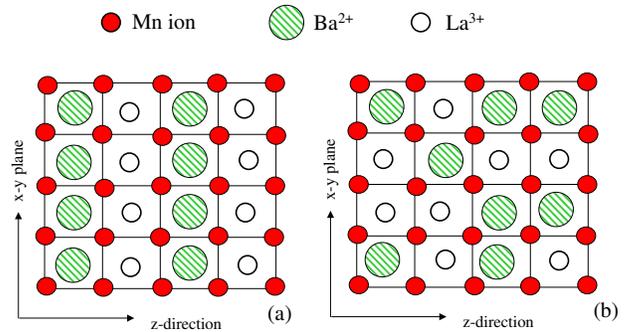}
  \caption{($color$ $online$) Schematic representation of an ordered (a) and disordered (b)
   La$_{0.5}$Ba$_{0.5}$MnO$_3$ perovskite. In the ordered case planes containing La and Ba cations
   alternate along the $z$-direction. In the disorder case the cations are randomly distributed.}
 \label{Scheme}
\end{figure}

Attfield and cooworkers\cite{Williams_03,Rodriguez-Martinez_96} found that in FM perovskites of the form
AMnO$_3$, the variation of Curie temperature with disorder is related with the distribution of the A cations
radius, $r_A$, and there is not significant dependence on the A site charge variance. Therefore, we expect that
the observed\cite{Sato_04} difference in $T_c$ between the ordered and disordered
La$_{0.5}$Ba$_{0.5}$MnO$_3$ perovskites is related with the different  distributions  of $r_A$. Note that
standard ionic radii values are $r_A = 1.21$ for La$^{3+}$ and $r_A = 1.47$ for
Ba$^{2+}$\cite{Rodriguez-Martinez_96}. In the ordered La$_{0.5}$Ba$_{0.5}$MnO$_3$ perovskite,
Fig.\ref{Scheme}(a), alternating planes of LaO and BaO separate  MnO$_2$ sheets in such a way that each Mn  ion
has four La$^{3+}$ neighbors in a direction  and four Ba$^{2+}$ neighbors in the opposite direction being  all
the Mn ions  equivalent. In the disorder perovskites, Fig.\ref{Scheme}(b), the La and Ba cations are randomly
distributed and the Mn's are surrounded by different combinations of divalent and trivalent cations. These
different chemical environments  traduce in different \textit{chemical pressures} on the Mn ions, and is a
source of disorder in the Mn electronic levels.

In this work we combine Ginzburg-Landau formalism with realistic two
orbital DE microscopic calculations, to study the effect that
the disorder in the position of the trivalent and divalent cations
has on the magnetic properties of FM manganites. The main results of
the work are, i) The Curie temperature, $T_c$ decreases with the disorder,
ii) as the disorder increases, the FM-PM phase transition becomes
more abrupt and eventually, for a  critical disorder, the magnetic
transition becomes first order and iii) we identify the variance of
disorder distribution as the relevant parameter for the decrease of
$T_c$. All these three points agree with the
experimental results\cite{Rodriguez-Martinez_96,Sato_04,Fontcuberta_98}.  Our
analytical results provide a framework for understanding  the
dependence of $T_c$ on disorder, both, in
experiments and in Monte Carlo simulations of simplified
models\cite{Motome_03,Kumar_03,Sen_04,Kumar_06}.

\textit{Model.} In a manganite of formula R$_{1-x}$A$_x$MnO$_3$,
there are $4-x$ electrons per Mn ion.
The crystal field splits the Mn $d$ levels
into an occupied strongly localized
$t_{2g}$ triplet an a doublet of $e_g$ symmetry. Furthermore, the
Hund's coupling in Mn ions is very large and aligns the spins of the
$d$ orbitals. Effectively, there are $1-x$ electrons per
Mn ions hoping between the empty $e_g$ Mn states.

The large Hund's coupling forces  each electron spin to align
locally with the core spin texture.
The spin of the carriers is conserved in the hoping between Mn ions, being the tunnelling amplitude maximum and
the kinetic energy  minimum,  when the spins of the Mn ions are parallel and the system is ferromagnetic. This
is basically the so called double exchange (DE) mechanism proposed fifty years ago by Zener\cite{zener} to
explain the FM order in manganites. When the temperature increases the kinetic energy minimized by the FM
order competes with the orientational entropy of the Mn core spin and at $T_c$, the
system becomes paramagnetic (PM)\cite{mjc}. In the PM phase the reduction of the kinetic energy could favor that
localization effects become more effective and a metal insulator transition could occur near the FM-PM
transition\cite{JAV}. In the case of La$_{0.5}$Ba$_{0.5}$MnO$_3$, the localization effects are rather weak and
the material is metallic at both sizes of the FM-PM transition. Therefore, in our model we do not consider
coupling with the lattice, however as the DE mechanism  depends strongly in the kinetic energy gain, we treat
the motion of the carriers in a realistic way by including the two $e_g$ orbitals.

\textit{Hamiltonian.} With the above considerations, the two orbital
DE Hamiltonian takes the following form,
\begin{equation}
H=\sum_{<i,j>,a,b}\, \,  f_{i,j}\, \,   t^{u}_{a,b}\, \, C^+_{i ,a}C^{ }_{j, b}\sum _{i,a} \epsilon_i \, \, \hat{n} _{i,a}  \, \, \, , \label{H}
\end{equation}
here $C^+_{i ,a}$ creates an electron in the Mn ion located at
site $i$ in the $e_g$ orbital $a$ ($a$=1,2 $1$=$|x^2-y^2>$ and
$2$=$|3z^2-r^2>$). In the limit of infinite Hund's coupling, the
spin of the carrier should be parallel to the Mn core spin ${\bf
S}_i$, and the tunnelling amplitude  is modulated by the spin
reduction factor
\begin{equation}
f_{i,j}= \cos\frac{\vartheta
_i}{2}\cos\frac{\vartheta _j}{2} + e^{i ( \phi _i - \phi _j)}
\sin\frac{\vartheta_i}{2}\sin\frac{\vartheta _j}{2} \label{srf}
\end{equation}
where $\{ \vartheta _i, \phi_i \}$ are the Euler angles of the Mn
core spins. The hopping amplitude depends both on the direction $u$
between sites $i$ and $j$ and the orbitals involved;
$t_{1,1}^{x(y)}=\pm \sqrt{3} t_{1,2}^{x(y)} =\pm \sqrt{3}
t_{2,1}^{x(y)}=3t_{2,2}^{x(y)}=t$. In the z direction the only
nonzero term is $t_{2,2}^{z}=4/3t$ \cite{dagottobook}. Hereafter $t$
is taken as the energy unit.

The last term describes the diagonal disorder.
$\hat{n} _{i,a} = C^+ _{i,a} C _{i,a} $ is the occupation
operator of orbital $a$ at site $i$ and $\epsilon _i$ is the energy
shift produced by the \textit{chemical pressure} on the Mn ion at
site $i$. This shift affects equally both $e_g$ orbitals. As
discussed above, the \textit{chemical pressure} at site $i$ depends
on the ionic radii of the cations surrounding the Mn ion. We assume
that the total shift at site $i$ is the sum of the shifts produced
by the eight next neighbors cations. Cations with big ionic
radius induce a positive energy shift $\Delta $ whereas cations
with small ionic radius induce a negative energy shift $-\Delta $.
With this election a Mn ion surrounded by equal number of small and
large cations has zero energy shift. In the case of a manganite of
composition R$_{0.5}$A$_{0.5}$MnO$_3$ the diagonal shift  takes
values in the range $-8 \Delta < \epsilon _i < 8 \Delta$. In this
description of the cationic disorder the strength of the disorder is
defined by the value of $\Delta$.

\textit{Free energy.} In order to describe thermal effects we have to compute the free energy. In manganites the
Fermi energy of the carriers is much greater than the typical Curie temperature and we only consider the
entropy of the classical Mn core spins. In the mean field approximation\cite{brey_2005} and in the limit of
small magnetization,  the entropy  per Mn core spin takes the form,
\begin{equation}
S(m)= \frac {\log 2}{2} - \frac {3}{2} m ^2 - \frac {9}{20} m ^4 + ...\, \, \label{entropy}
\end{equation}
where $m$ is the thermal average of the relative magnetization of the Mn core spins.

At finite temperatures the magnetization is not saturated and the
spin reduction factor, Eq. (\ref{srf}) is smaller than unity. Treating
the spin fluctuations in the virtual crystal approximation, the spin
reduction factor  is substituted by its expectation value, that for
small magnetization has the form\cite{DeGennes},
\begin{equation}
f_{i,j} \simeq {f}_{m} = \frac {2}{3} - \frac {2}{5} m ^2 - \frac{6}{175} m ^2+ ... \, \,
\label{fm}
\end{equation}
With this the thermal average of the internal energy per Mn ion can be written in  the form,
\begin{eqnarray}
E &   = \! \!  \frac {{f}_m}{N}  \left ( \! \sum t^{u}_{a,b} < \! C^+_{i ,a}C^{ }_{j, b} \!> \!   +  \! \frac {1}
{{f}_m} \sum \epsilon_i < \!    \hat{n} _{i,a}        \!>  \! \right ) \label{Internal_energy}
\end{eqnarray}
being $N$ the number of Mn ions in the system. In
Eq. (\ref{Internal_energy}), it is evident that the relative importance
of the disorder increases when the magnetization, and thus the spin
reduction factor decreases. For small values of $\Delta$ the
internal energy can be expanded in powers of the disorder strength.
As the mean value of $\epsilon _i$  is zero, the first term
different from zero is proportional to the variance, $\sigma ^2
(\epsilon _i)$, of the diagonal disorder distribution,
\begin{equation}
E= f _m  \left ( E _0 + a \frac {\sigma ^2 (\epsilon _i)}{{f} ^2 _m} \right ) \label{Internal1} \, \, \, ,
\end{equation}
where $E_0 < 0 $ is the kinetic energy per Mn ions in the disorder free FM phase, and
\begin{equation}
\sigma ^2 (\epsilon _i) = \frac {1}{N} \sum _ i  { \epsilon _i }^2
\label{variance} \, \, \, .
\end{equation}
In Eq. (\ref{Internal1}), the coefficient $a$ is negative as the
electrons prefer to place on sites with negative values of
$\epsilon_i$.

Combining Eq. (\ref{entropy}-\ref{variance}), the dependence of the   free energy, $F=E-TS$, on the magnetization,
can be written as
\begin{eqnarray}
\! \! \! \! \! \! \! \! \! & F  & =     \frac {3}{2} T m ^2 + \frac {2}{5}  \left ( E _0 \! -  \! \frac{9}{4} \,
a \, \sigma ^2 (\epsilon _i) \right )\, m ^2  \nonumber \\
 &  & \! \! \!
+ \frac {9}{20} T m ^4  \! - \! \frac {6}{175} \left ( E _0 \! - \! \frac{39}{2} \, a \, \sigma ^2 (\epsilon _i)
\right )\, m ^4 \, + ... \, \, \, \, \, \label{freeenergy}
\end{eqnarray}
and the Curie temperature takes the form
\begin{equation}
T_c = - \frac {4}{15} \left  ( E _0 \! -  \! \frac{9}{4} \, a \,
\sigma ^2 (\epsilon _i) \right ) \, \, \, , \label{tc}
\end{equation}
where is clear that $T_c$ decreases when the disorder  strength  increases. The order of the
transition can be inferred from the sign of the quartic term in Eq. (\ref{freeenergy}). If the quartic term is
positive the transition is second order, while a negative quartic term implies the existence of a first order
phase transition. Using the previous expression for $T_c$, a first order transition takes place if,
\begin{equation}
\sigma ^2 (\epsilon _i)
>   \frac {2}{39} \frac {E_0}{ a} \, \, \left ( \frac {1+ \frac {7}{2}}{1 + \frac {21}{52}} \right
) \label{firstorder} \, \, \, .
\end{equation}
Equations (\ref{tc}) and (\ref{firstorder}) are the main results  of
this work, they show that the Curie temperature decreases when the
strength of the disorder increases and that for disorder strong
enough the FM-PM transition  changes from second to first order.
Comparing Eq. (\ref{tc}) and Eq.(\ref{firstorder}) it results that for
changes in the Curie temperature larger than 10 per cent, the FM-PM  transition
transforms from second to first order.

\textit{Numerical results.} In order to check the approximations we
have done  and the validity of the obtained results, we diagonalize
numerically the ferromagnetic two orbital DE Hamiltonian,
 in presence of diagonal disorder, Eq. (\ref{H}). We
consider the case of disorder in the ionic radii of the cations
surrounding the Mn ions. As discussed above, this disorder produces a
distribution of diagonal energy shifts in the range $-8 \Delta <
\epsilon _i < 8 \Delta$. For  a random distribution of the two type
of cations, the variance of the diagonal disorder distribution is
$\sigma ^2 (\epsilon _i) = 8 \Delta ^2$. We have diagonalized the
Hamiltonian for different disorder realization and different values
of the disorder strength $\Delta$. We work with a cluster containing
12$\times$12$\times$12 atoms and with periodic boundary conditions.
By studying smaller clusters we have checked that our results are
free of finite size effects.

\begin{figure}
  \includegraphics[clip,width=8cm]{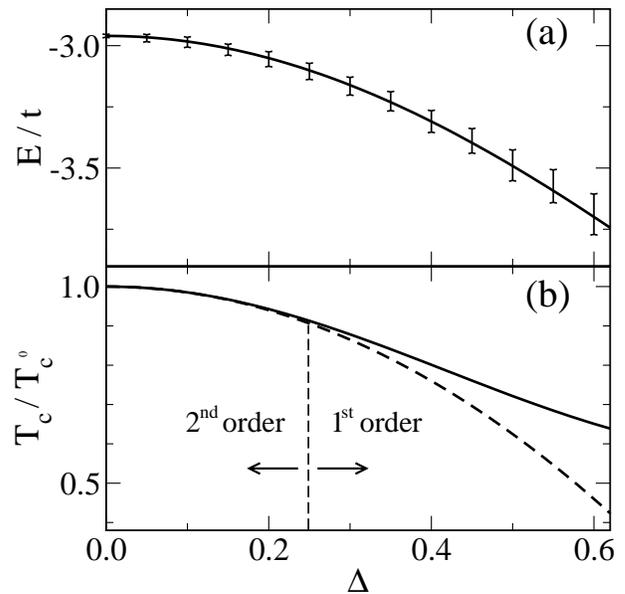}
  \caption{ Polynomial fit to a) internal  energy  and b) Curie temperature
   as obtained  from numerical simulations. Bars in a) indicates
  the typical indetermination in the numerical calculations.
In b) the dashed line corresponds to the  quadratic approximation to
$T_c$ (Eq. (\ref{tc})).  In b) the regions with first and second
order  FM-PM  transition are shown.}
 \label{fig1}
\end{figure}

In Fig.\ref{fig1}a we plot, for the full polarized FM phase, the
internal  energy per Mn ion  as function of the disorder strength
$\Delta$. As expected\cite{Mahan_book}, for small values of the
disorder the internal energy decreases quadratically with $\Delta$.
Taking into account the dependence of the internal energy on the
spin reduction factor, in mean field approximation the Curie
temperature gets the form
\begin{equation}
T_c = - \frac {2}{3} \frac {\partial \, U} {\partial \, m ^2} \, \,
\, .\label{Tc2}
\end{equation}
In Fig.\ref{fig1}b, we plot $T_c$, as obtained numerically
from the internal energy, as function of $\Delta$. In the disorder free case,
 $T_c \sim 0.8 t$ that compared  with the
Curie temperature of the ordered La$_{0.5}$Ba$_{0.5}$MnO$_3$,
$T_c$=350K, \cite{Sato_04}, implies a value of the hopping amplitude
of $t \sim 0.038eV$.  For changes up to 20 per cent, we find that
the decrease of $T_c$ with $\Delta$ can
be fitted very accuracy with a quadratic dependence. Experimentally
the differences in $T_c$ between the ordered and disordered samples
in La$_{0.5}$Ba$_{0.5}$MnO$_3$ is near  14 per cent\cite{Sato_04},
therefore we expect that for manganites with atomic radii size disorder,
the expressions obtained analytically for $T_c$, Eq. (\ref{tc}),
and the order of the transition,
Eq. (\ref{firstorder})  should be valid.

In fig.\ref{fig1}b, we also indicate the disorder strength for which
the magnetic transition changes form second to first order. As
discussed above, the transition becomes first order for changes in
$T_c$ larger than 10 per cent. In the case of
La$_{0.5}$Ba$_{0.5}$Mn O$_3$, the disorder produces changes in $T_c$
 of near 14 per cent and therefore our results
explain  the experimental observed change in the order of the
magnetic transition\cite{Sato_04}.  We expect that in FM
manganites  with smaller differences in the atomic radii than in
La$_{0.5}$Ba$_{0.5}$MnO$_3$, for example La$_{0.5}$Sr$_{0.5}$MnO$_3$
($r_A = 1.21$ for La$^{3+}$ and $r_A = 1.31$ for
Sr$^{2+}$\cite{Rodriguez-Martinez_96}) the effect of the disorder in
 $T_c$  should be smaller and the order of the
magnetic transition should be second order independently of the
order in the position of the cations.

Finally we want to check that, as proposed by
Rodr\'{\i}guez-Mart\'{\i}nez and
Attfield\cite{Rodriguez-Martinez_96}, the relevant figure for
quantifying  the disorder is the variance of the disorder
distribution.
We have repeated the numerical calculation of $T_c$
for different distribution of diagonal disorder. Apart from the
\emph{cation disorder} distribution , we have also analyzed a
\emph{uniform }distribution of the diagonal energy shift between two
values,  a \emph{binary} distribution and  a \emph{gaussian}
distribution of the diagonal disorder.

\begin{figure}
  \includegraphics[clip,width=8cm]{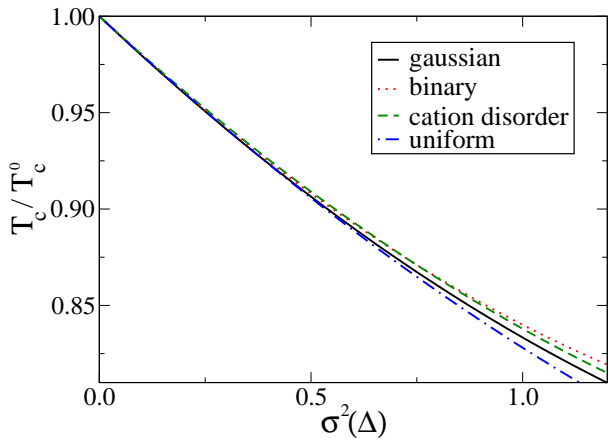}
  \caption{($color$ $online$) Curie temperature versus  variance of diagonal disorder
   for different models of disorder.  The different kinds of disorder are explained in the text}
 \label{fig2}
\end{figure}

The results are plottted in Fig.\ref{fig2}. For all models studied,
we find that for moderate disorder strength, the Curie temperature
decreases linearly with the variance of the disorder. There are
deviations from this dependence in the limit of strong disorder. In
this limit  the carriers start to localize in the sites with smaller
energy and perturbation theory becomes not valid. This regime is not
the relevant one in FM manganites, and the results shown in Fig.
\ref{fig2} agree with
the experimental results obtained by Rodr\'{\i}guez-Mart\'{\i}nez
and Attfield\cite{Rodriguez-Martinez_96}.
The numerical results also support our analytic findings that for moderate disorder
strength the variation of the Curie temperature and the order of the
transition depend on the the variance of the distribution of the
diagonal energy shifts in the electronic Hamiltonian.

In summary, we have studied a realistic model of manganites by means
of exact microscopic calculations and Landau Theory formalism. The
different sizes of cations surrounding a manganese ion are included
in the model as an energy shift in that manganese site. This model
reproduces the experimental results. It explains the observed strong
reduction of $T_c$ in disordered samples with respect to ordered
ones\cite{Sato_04}. Moreover, it demonstrates, in agreement with
experiments\cite{Sato_04}, that disorder makes the FM-PM transition
more abrupt and, for a enough disorder strength, this transition
becomes first order. Finally our formalism  identifies in a natural
way the variance of the distribution of the diagonal energy shifts,
as the relevant parameter to characterize changes in the magnetic
properties, independently of the model of
disorder\cite{Rodriguez-Martinez_96, Fontcuberta_98,Collado}.


\vspace{0.5truecm}

{\it Acknowledgements.} Financial support is acknowledged from Grant
F.P.U. Ref. AP-2004-0772 (MEC, Spain).


\end{document}